\documentclass[11pt]{elsarticle}

\usepackage{amsmath,amssymb,amsthm}
\usepackage{physics}
\usepackage{graphicx}
\usepackage{wrapfig}
\usepackage{float}
\usepackage{hyperref}
\usepackage{geometry}
\usepackage{subcaption}
\usepackage{enumitem}
\hypersetup{
    colorlinks=true,
    linkcolor=blue,
    citecolor=blue,
    urlcolor=blue
}

\title{Variational Quantum Brushes}
\begin{document}

\author{Jui-Ting Lu}
\address{Université de Lorraine, CNRS, LORIA, Nancy, France.}

\author{Henrique Ennes}
\address{Inria d'Université Côte d'Azur, Sophia Antipolis, France.}

\author{Chih-Kang Huang}
\address{Universit\'{e} de Lorraine, CNRS, IJL, Nancy, France.}
\address{Universit\'{e} de Lorraine, CNRS, Inria, IECL, Nancy, France.}

\author{Ali Abbassi}
\address{LIST3N, Université de Technologie de Troyes, Troyes, France.}
\address{Orange Research, Châtillon, France.}

\begin{abstract}
Quantum brushes are computational arts software introduced by \emph{Ferreira et al} (2025) that leverage quantum behavior to generate novel artistic effects.
In this outreach paper, we introduce the mathematical framework and describe the implementation of two quantum brushes based on variational quantum algorithms, Steerable and Chemical.
While Steerable uses quantum geometric control theory to merge two works of art, Chemical mimics variational eigensolvers for estimating molecular ground energies to evolve colors on an underlying canvas.
The implementation of both brushes is available open-source at \url{https://github.com/moth-quantum/QuantumBrush} and is fully compatible with the original quantum brushes.
\end{abstract}

\maketitle

\begin{figure}[H]
    \centering
    % \foreach \file in {fig/Renoir8/Bal_du_molin_de_la_Galette_c2_t0p00.png,
    %  fig/Renoir8/Bal_du_molin_de_la_Galette_c2_t0p14.png,
    %  fig/Renoir8/Bal_du_molin_de_la_Galette_c2_t0p29.png,
    %  fig/Renoir8/Bal_du_molin_de_la_Galette_c2_t0p43.png,
    %  fig/Renoir8/Bal_du_molin_de_la_Galette_c2_t0p57.png,
    %  fig/Renoir8/Bal_du_molin_de_la_Galette_c2_t0p71.png,
    %  fig/Renoir8/Bal_du_molin_de_la_Galette_c2_t0p86.png,
    %  fig/Renoir8/Bal_du_molin_de_la_Galette_c2_t1p00.png
    % }{%
    %     \includegraphics[width=0.22\linewidth]{\file}
    % }
    \includegraphics[width=0.22\linewidth]{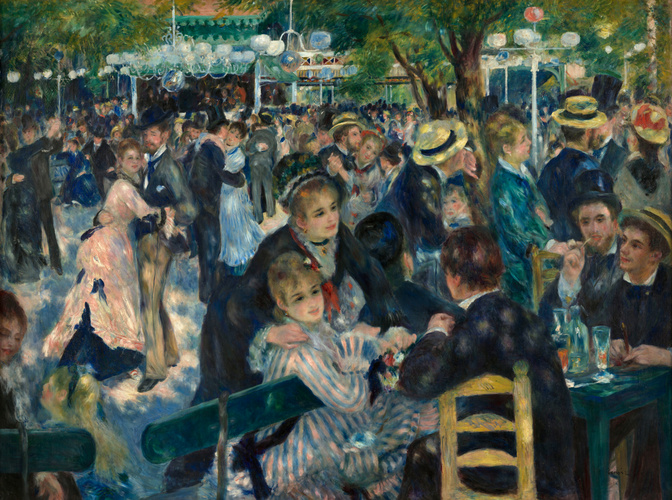}
    \includegraphics[width=0.22\linewidth]{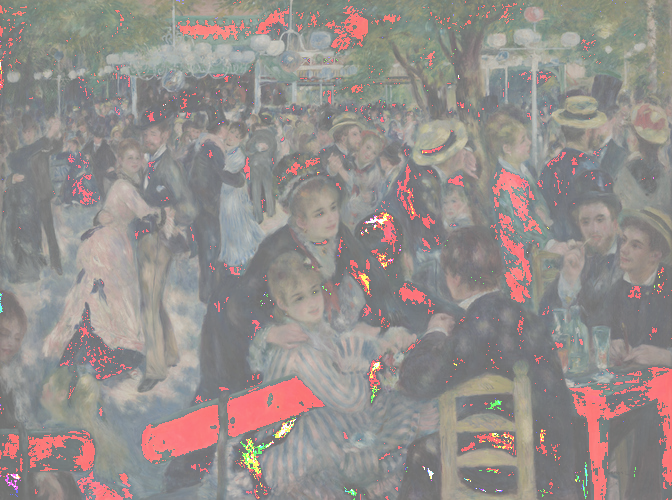}
    \includegraphics[width=0.22\linewidth]{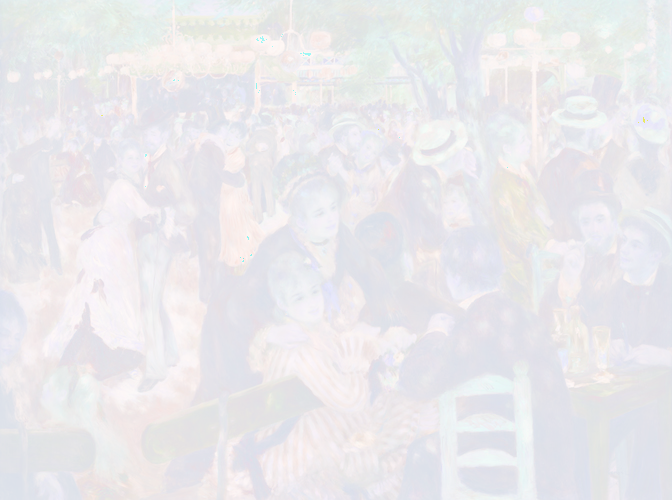}
    \includegraphics[width=0.22\linewidth]{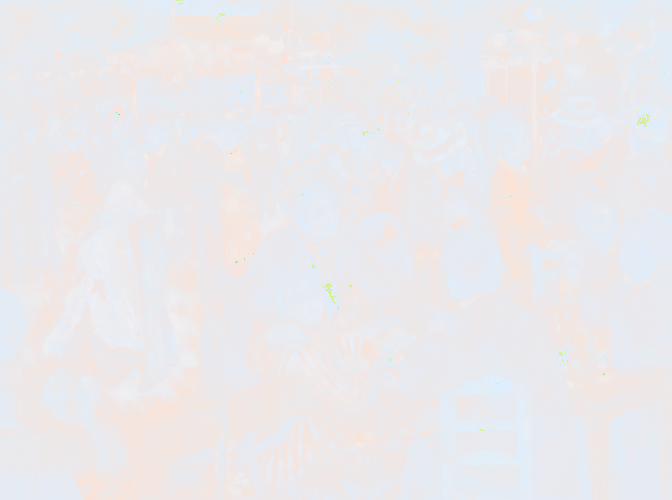}
    \includegraphics[width=0.22\linewidth]{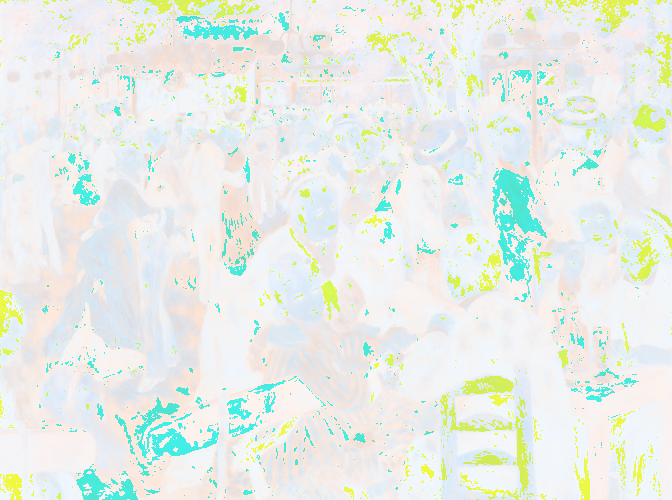}
    \includegraphics[width=0.22\linewidth]{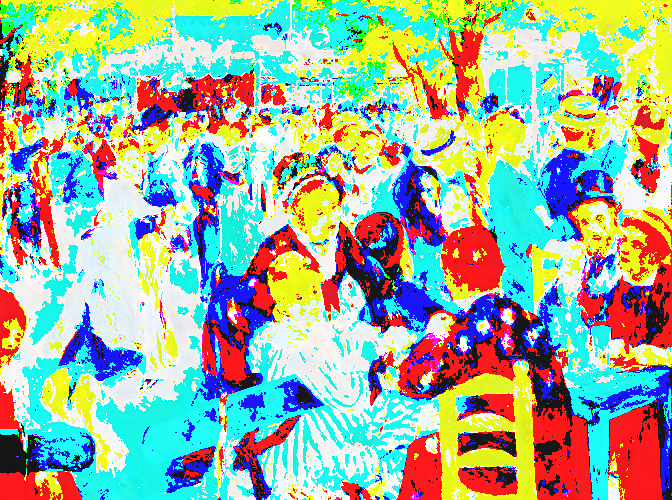}
    \includegraphics[width=0.22\linewidth]{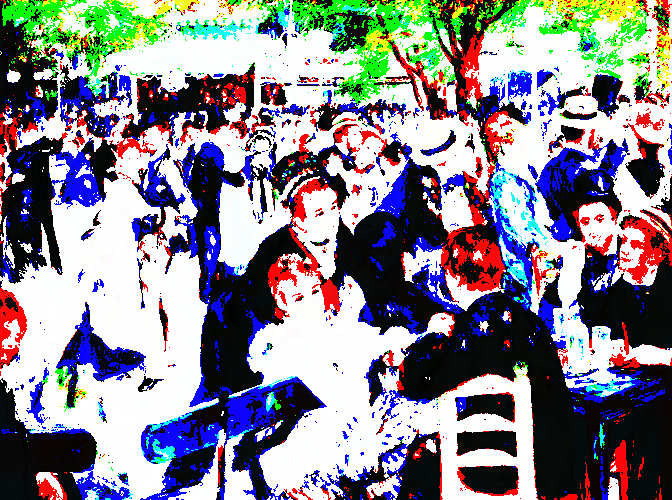}
    \includegraphics[width=0.22\linewidth]{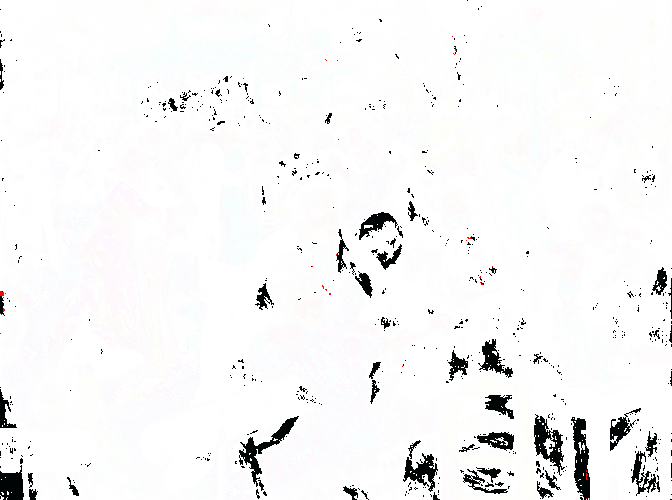}
    \caption{Steerable effect applied to Renoir's \emph{Bal du moulin de la Galette} using a red parrot as target image.}\label{fig: renoir}
\end{figure}

\section{Introduction}
\emph{Variational quantum algorithms} (\emph{VQA}s) encompass some of the most anticipated techniques for real quantum advantage over digital computing. 
Originally inspired by applications in computational chemistry \cite{mcclean2016theory, peruzzo2014variational} -- where the minimization of molecular energy remains an important open question -- these methods were soon extended to more general optimization frameworks, ranging from combinatorial problems \cite{cerezo2021variational, farhi2014quantum, farhi2016quantum, lucas2014ising}, to portfolio optimization tasks \cite{agliardi2025portfolioconstructionusingsamplingbased}. 
Abstractly, all these approaches share the iterative use of classical, typically gradient-based, optimization techniques to minimize objective functions evaluated on parametrized quantum states, therefore mirroring the structure of classical variational methods common in machine learning \cite{bishop2006pattern, nakajima2019variational}.
Since quantum Hilbert spaces can encode information that grows exponentially with the number of qubits, such methods are expected to enable efficient solutions to problems that are believed to be intractable for conventional algorithms.
On the more practical side, VQAs belong to the very restricted classes of methods known to be noisy intermediate-scale quantum (NISQ) \cite{RevModPhys.94.015004}. 
Coined by Preskill in 2018 \cite{preskill2018quantum}, this term is used to designate quantum hardware particularly susceptible to noise and with few computational qubits, exactly the kind of machine currently available. 
For this reason, more practical applications of VQAs are already abundant; particularly interesting to us is their use in quantum arts, mainly in quantum music \cite{itaborai2023variational, Miranda_2024}.

NISQ is also the main computational assumption of the \emph{quantum brushes} introduced by \cite{ferreira2025quantumbrush}, a widely accessible quantum-driven software for the visual arts. 
In a nutshell, these brushes appeal to quantum effects -- such as the no-cloning theorem and entanglement -- to describe the local evolution of colors on an underlying canvas, creating novel artistic expressions for practitioners of computational arts.
By restricting their brushes to NISQ-only effects, Ferreira \emph{et al} could not only classically simulate the quantum effects in digital computers but also run them on real quantum hardware, with noise playing a surprising role in the final visual output. 
These open-source brushes have already been used to generate fascinating works of art and form an exciting direction for both the popularization and the practical use of quantum computing techniques to an audience beyond the usual researchers of the field.

Here we take the next natural step by describing the implementation of two brushes based on variational quantum algorithms, \emph{Steerable} and \emph{Chemical}. 
Although both are based on variational methods, their construction and artistic aims are significantly different.
{Steerable} uses concepts of geometric control theory to merge two pieces of art, represented as distinct quantum states.
Simply put, geometric control theory studies how systems governed by physical laws evolve when their motion can be influenced by certain controllable degrees of freedom.
It tells us, for example, that a car can park in parallel, even though it cannot move directly sideways.
For us, control theory will be used in its quantum incarnation to smoothly drive a region of an input canvas into another. 
This can be visualized as a \emph{path} of canvases whose beginning is full of pieces very similar to the original image, but that start to look more and more like the target as one gets to its end.
The user, by controlling a continuous parameter, decides where to stop on this road. 

Chemical, on the other hand, uses the minimization procedure of \emph{variational quantum eigensolvers} (\emph{VQE}) to generate a parametric family of quantum circuits that, at the convergence limit to optimal parameters, describes the ground state of some molecules.
This means that while the variational parameters change throughout the optimization, the \emph{physical process} dictated by them gets closer and closer to the evolution of a fixed ansatz state towards the minimal energy configuration.
By applying these circuits to the angular encoding of different pixels in a finite stroke, we define a path of states whose final points evolve like a molecular system, while their intermediate values "record" the steps of the variational algorithm.
Therefore, not only will the brush be driven by a real-life chemical process but also by the algorithmic tool used in current quantum research to \textit{study} that same process.
In a sense, it is an attempt to bring cutting-edge scientific evolution to the artists' fingertips.

The objectives of this paper are two-fold. 
First, it serves as a technical report of the implementations\footnote{The code is fully available online. The two variational brushes are incorporated into the original \emph{QuantumBrush} application and can be installed from \url{https://github.com/moth-quantum/QuantumBrush}. Extra implementation details are given at \url{https://github.com/mothorchids/Luminists-Quantum-Brushes}.} of Steerable and Chemical, where we describe the mathematical framework, as well as details that did not fit the code comments.
In this sense, we try to follow the structure of \cite{ferreira2025quantumbrush}, in which the technical aspects of each brush are described separately (in Sections \ref{sec: steerable} and \ref{sec: chemical}, respectively) and, therefore, can be read independently.
Second, we hope that this paper fosters the dissemination of variational quantum algorithms to practitioners of quantum computing and enthusiasts of computational art. 
Therefore, we expect the reader to be acquainted with the fundamentals of quantum computing at the level of \cite{ferreira2025quantumbrush} -- for an accessible review, refer to \cite{montanaro2016quantum} -- but we introduce aditional theoretical background of variational methods throughout the text. 
We hope that our brushes correspond only to a first step towards the full incorporation of VQA methods into the visual arts.

\paragraph{Acknowledgments}
The two quantum brushes described in this paper were originally implemented as part of the authors' contribution to the \emph{Bradford Quantum Hackathon 2025}, organized jointly by \emph{Quantinuum} and \emph{Aqora}.
We are grateful to the hackathon organizers for welcoming us, especially Astryd Park from \emph{MOTH Quantum} for valuable discussions and clarifications regarding the original quantum brushes.
We would also like to thank Massinissa Zenia, who participated in the initial phases of this project.

\section{Steerable}\label{sec: steerable}
In its quantum iteration, control theory is typically concerned with manipulating the behavior of quantum systems, such as atoms, molecules, or qubits, using precisely shaped external fields like lasers, microwaves, or magnetic pulses.
By fine-tuning these controls, scientists can steer a system’s evolution toward a desired quantum state or outcome. 
For example, laser pulses can be optimized to drive chemical reactions along specific pathways, or microwave signals can be used to perform accurate qubit rotations in a quantum computer. 
Through such techniques, quantum control\footnote{For an overview of quantum control theory, including its mathematical background, refer to \cite{d2021introduction}.} enables both fundamental discoveries in physics and practical advances in quantum technologies \cite{brif2010control, d2021introduction, glaser2015training}.

\emph{Steerable} applies this framework to drive one artwork into another.
Notice that, despite an intent similar to the Collage brush already implemented by \cite{ferreira2025quantumbrush}, Steerable is mechanically very different.
Instead of attempting to copy the {representative quantum} state $\ket{p_0}$ of the source {work}, Steerable looks for a parametrized family of quantum circuits $U(t)$ such that, when $t=0$, it does not change $\ket{p_0}$, but when $t=1$, it reaches a state very close to $\ket{p_1}$, the state of the target canvas.
This means that for values of $t$ in between 0 and 1, {the evolved state} $U(t)\ket{p_0}$ will represent a canvas "in between" the original and the target, although \emph{not} a mixture of both -- or, at least, not in the classical sense of the word.
Quantum control, therefore, opens an entirely new creative frontier. 
Through controlled quantum dynamics, one can generate visual and interactive art that reflects the inherent beauty and unpredictability of quantum mechanics, translating the invisible behaviors of the quantum world into tangible aesthetic experiences.

\subsection{Theoretical framework}\label{subsec:theoretical_steerable}
Suppose that the two canvases have been discretized into pixel grids and that $\mathcal{C}_1$ and $\mathcal{C}_2$ are lasso-defined copy and target regions.
We can further assume that $\mathcal{C}_1$ and $\mathcal{C}_2$ are represented by quantum states, respectively denoted $|p_0\rangle$ and $|p_1\rangle$.
We aim to steer $ |p_0 \rangle$ to $|p_1 \rangle$ using quantum circuits representing \emph{drifts} and \emph{controls} through a path $|\rho(t)\rangle$ in the common Hilbert space of the least total energy possible\footnote{The reader familiar with the mathematical formulation of classical mechanics will notice that this path corresponds to the minimization of an energy functional.}. 
This can be modeled by a \emph{bilinear multilevel} quantum system \cite{albertini2003notions}, which amounts to solving the following optimization problem
\begin{equation}
\boldsymbol{\eta}^* = \arg\min_{\boldsymbol{\eta}} \left( 1 - F(|\rho(1)\rangle, |p_1\rangle) + \int_0^1 \sum_{i=1}^m u_i(t;\boldsymbol{\eta})^2 , dt \right),
\label{eq:opt_control}
\end{equation}
subject to
\begin{equation}
\begin{cases}
\frac{d}{dt} |\rho(t)\rangle = - i H(t;\boldsymbol{\eta}) |\rho(t)\rangle, \\
|\rho(0)\rangle = |p_0\rangle,
\end{cases}
\label{eq:QNODE}
\end{equation}
where $F(|\rho(1)\rangle, |p_1\rangle)$ is the quantum fidelity between $|\rho(1)\rangle$ and $| p_1 \rangle$, a real number in the interval $[0,1]$ which equals $1$ if and only if $|\rho(1)\rangle$ equals $ |p_1\rangle$ up to a global phase. 
The Hamiltonian is given by
\begin{equation}
H(t;\boldsymbol{\eta}) = H_0 +  u_1(t;\boldsymbol{\eta}) H_1 + \ldots u_m(t;\boldsymbol{\eta}) H_m,
\label{eq:multilinear_hamiltonians}
\end{equation}
where $H_0$ is the \emph{drift} Hamiltonian, $H_1, \ldots, H_m$ are the \emph{control} Hamiltonians, and the real functions $u_1(t;\boldsymbol{\eta}), \ldots, u_m(t;\boldsymbol{\eta})$ are the \emph{control amplitudes} parameterized by the \emph{optimization variable} $\boldsymbol{\eta}$.
Note that the control amplitudes are the only adjustable parameters in this optimization problem.
We see that the two terms of equation~\eqref{eq:opt_control} naturally correspond to the two constraints of the path $|\rho(t)\rangle$; that is, 1. it ends at $|p_1\rangle$, and 2. it is the path of minimum energy.
For the sake of simplicity, we will omit the explicit dependence on $\boldsymbol{\eta}$ in the following and simply write $H(t)$ and $u_i(t)$.

A choice of Hamiltonians $H_0, H_1,\dots,H_m$ for which there always exists some control amplitudes $\boldsymbol{u}(t)= (u_1(t),\dots,u_m(t))$ for any source $|p_0\rangle$ and target $|p_1\rangle$ that solves~\eqref{eq:QNODE} is said to define an \emph{exactly controllable} system.
Exactly controllability has been extensively characterized in the literature~\cite{albertini2003notions, ramakrishna1995controllability} and we assume, without a proof\footnote{The easiest way to see that this is indeed the case is to note that the choice of Hamiltonians in equation~\eqref{eq:control_hamiltonians} spans a Lie algebra isomorphic to $\mathfrak{su}(2^n)$, where $n$ is the assumed number of qubits \cite{albertini2003notions}. In fact, exact controllability of the system \eqref{eq:QNODE} can be achieved with a drift Hamiltonian $H_0$, and only one control Hamiltonian $H_1$, see \cite{kuranishi1951everywhere,turinici2000controllability}. However, such Hamiltonians consist of large sums of Pauli operators, which makes them impractical from a quantum control perspective. Exact controllability with a more natural set of Hamiltonians, such as tensor products of Pauli operators (Pauli strings), has been investigated in \cite{debnath2016demonstration,smith2025optimally}.}, that all systems discussed in this paper are exactly controllable.
Moreover, we shall assume $\ket{p_0}$ and $\ket{p_1}$ to be states of $n$ qubits, which implies that $H_0,H_1,\dots,H_m$ can be represented by $2^n\times 2^n$ complex Hermitian matrices.
In particular, if $H(t)$ is any linear combination of the drift and control Hamiltonians, $U = \exp(-iH(t))$ can be expressed by quantum gates.
For simplicity, we will always suppose that the number of controls, $m$, equals the number of available qubits, $n$. 

Once the appropriate Hamiltonians and Hilbert spaces are chosen, 
the remaining challenge is to determine explicit controls $\boldsymbol{u}(t)$ capable of steering $\ket{p_0}$ to the target state $\ket{p_1}$. 
Because of the time dependence of the Hamiltonian, it is in general difficult to obtain analytical solutions to \eqref{eq:QNODE}, forcing the use of numerical techniques. 
Several algorithms have been proposed for that, such as GRAPE \cite{khaneja2005optimal} and CRAB \cite{caneva2011chopped}.
Here, we opt to approximate the controls using neural networks, which, to the best of our knowledge, remain a relatively unexplored option. 
Besides being easy to implement, neural networks offer a powerful analytical framework for studying properties such as stability in architecture design, see for example~\cite{cao2025quantum}.

Explicitly, we train a neural network to approximate $u(t)$ by minimizing \eqref{eq:opt_control} and optimizing the parameters $\boldsymbol{\eta}$ through backpropagation.
To evaluate $\ket{\rho(1)}$ in \eqref{eq:opt_control}, we first note that
\begin{equation}\label{eq: long}
\begin{split}
     |\rho(1)\rangle & = U(1) \ket{\rho(0)}\\
     &=\mathcal{T}\exp\bigg(-i\int_0^1 H(t)\, dt\bigg) | \rho(0)\rangle\\
     &=\mathcal{T}\exp\bigg(-i\int_0^1  H_0 + u_1(t)H_1 + \ldots + u_m(t) H_m \, dt\bigg) | \rho(0)\rangle
\end{split}
\end{equation}
{where $\mathcal{T}$ is the time-ordering operator.}
Since the Hamiltonians do not commute, the exponential of the last of equation~\ref{eq: long} cannot be directly written as a product of exponentials\footnote{This is a result of the \emph{Baker–Campbell–Hausdorff formula}~\cite{nielsen2001quantum}.}.
One can express a parametric family of circuits for
$U(1)$ by approximating the time-ordered exponential using a second-order splitting scheme
\begin{equation}\label{eq:splitting circuit}
\begin{aligned}
U(1) \approx &\prod_{k=1}^N \left ( \exp(-i H_0 \frac{\Delta t}{2}) \exp(-i u_1(k \Delta t) H_1 \frac{\Delta t}{2}) \ldots 
\exp(-i u_m(k \Delta t) H_m \frac{\Delta t}{2}) \right. \\ 
&\quad \quad \left.
\exp(-i u_m(k \Delta t) H_m \frac{\Delta t}{2}) 
\ldots \exp(-i u_1(k \Delta t) H_1 \frac{\Delta t}{2}) \exp(-i H_0 \frac{\Delta t}{2})\right )
\end{aligned}
\end{equation}
with $N$ the number of discrete timesteps within the interval $[0,1]$ and $\Delta t = 1/N$.
Although complicated, the circuits described by \eqref{eq:splitting circuit} can be efficiently implemented as strings of Pauli gates, making the preparation of $\ket{\rho(t)}$ feasible (see Figure~\ref{fig:splitting_circuit_tikz}). 
In our test case, training is usually fast and achievable with shallow neural networks; however, we note that faster but less accurate approximations are possible through early stopping.
Furthermore, this training approach is unsupervised and not data-driven -- unlike most machine-learning applications -- 
which does not require tuning regularization parameters related to data overfitting.

Once $\boldsymbol{u}(t)$ is estimated, $\ket{\rho(t)}$ can be evaluated for any $t>0$ as in \eqref{eq:splitting circuit}. That is, while the constraints sets $\ket{\rho(0)}=\ket{p_0}$ and, for good solutions, $\ket{\rho(1)}$ is close $\ket{p_1}$, nothing prevents us from considering states $\ket{\rho(t)}$ for $t>1$.
This means that, instead of using {Steerable} only to interpolate between $\ket{p_0}$ and $\ket{p_1}$, we can also use it to \emph{extrapolate} to states beyond the target.
Using the analogy of a path of canvases, this can be seen as extending the road beyond its final stop, while still retaining its geometric information. 

It should be noted that, despite being hybrid and involving a minimization problem, the quantum control technique described here deviates significantly from the usual VQA approach.
In particular, no reference to the important variational principle~\cite{tilly2022variational} is necessary.
Nevertheless, Steerable bears enough similarities to variational methods that it deserves to be called a variational brush on its own.

\subsection{Implementation}\label{subsec:implementation_steerable}
\begin{figure}[H]
   \centering
    \includegraphics[width=\textwidth]{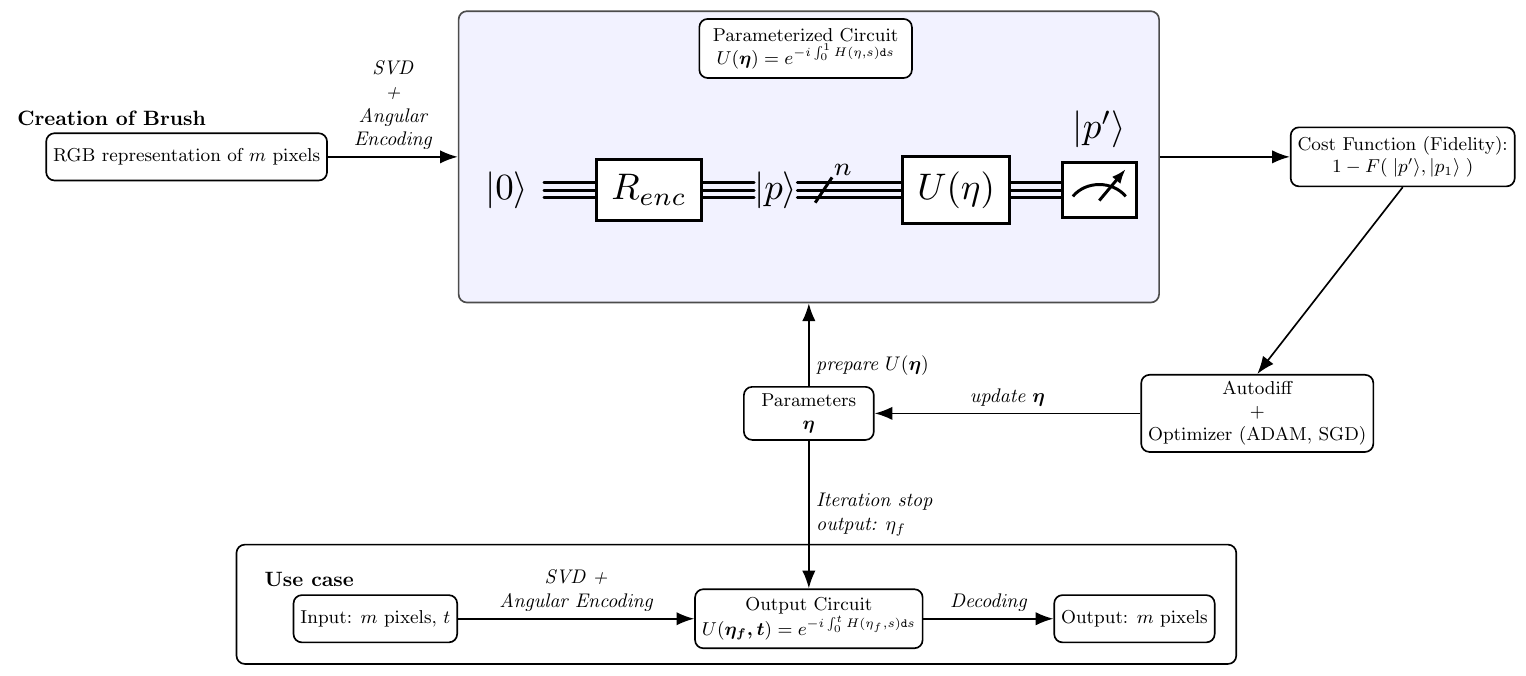}
    \caption{Diagram summarizing the learning process behind Steerable.}
    \label{fig:nice_diagram}
\end{figure}

The implementation of the steerable brush is summarized in the diagram in Figure~\ref{fig:nice_diagram}, which illustrates the flow of information throughout the system. 

We start by describing the user-defined parameters.
The user marks three regions -- source, target, and paste -- by directly sketching them on the canvas. 
The paste region may also be specified as a single point (i.e., two circles and one point), in which case the steered source image is pasted into the region whose barycenter corresponds to that point. 
By default, \texttt{"Source = Paste"} is \emph{false}; it should be enabled only when no separate paste region is defined.

A model is then trained to compute a quantum circuit that transforms the source region into the target region. 
After training, the learned transformation is applied to the paste region (which may be identical to the source).
The main parameter is \emph{\(t\)}, which controls how far along the learned transformation the system evolves:
\begin{align*}
t = 0 &\rightarrow \text{source-like},\\
t = 1 &\rightarrow \text{target-like},\\
t > 1 &\rightarrow \text{extrapolation beyond the target (may produce novel effects).}
\end{align*}

Additional parameters include \texttt{timestep}, that is number of discrete steps used to approximate the circuit’s evolution. 
It also controls the degree of transition smoothness in the brush dynamics.
Its default value is 25.
The other user-defined parameter, \texttt{controls}, sets the number of features used to represent a patch or, equivalently, the number of qubits in the circuit.
Optional visualization controls include \texttt{"show source \& target"} (along with boundary settings \texttt{"show color"} and 
\texttt{"show thickness"}), which displays the reference regions with boundaries rendered in the selected color and thickness.

\begin{figure}[ht]
    \centering
    \includegraphics[width=0.8\linewidth]{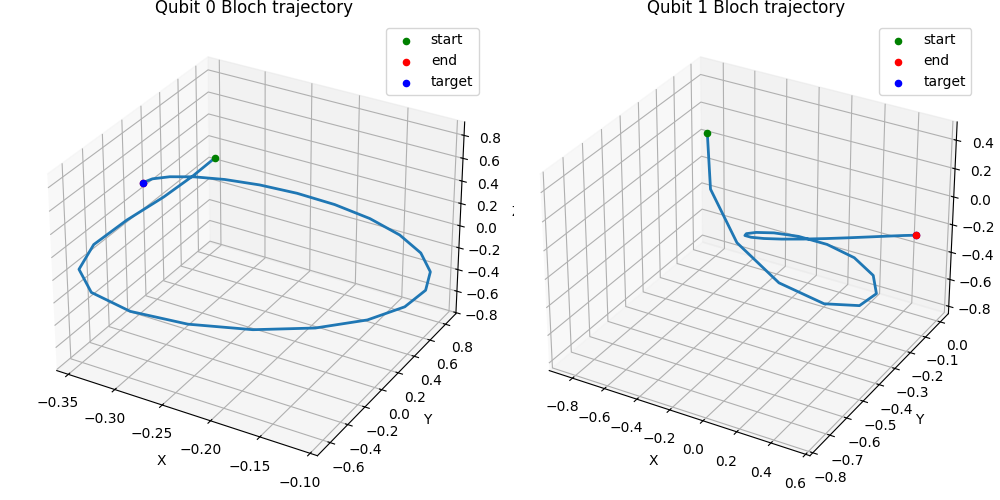}
    \caption{Visualization of the steering effect for 2 qubits.}
    \label{fig:steerable_NN}
\end{figure}

\begin{figure}[ht]
    \centering
    \includegraphics[width=0.8\linewidth]{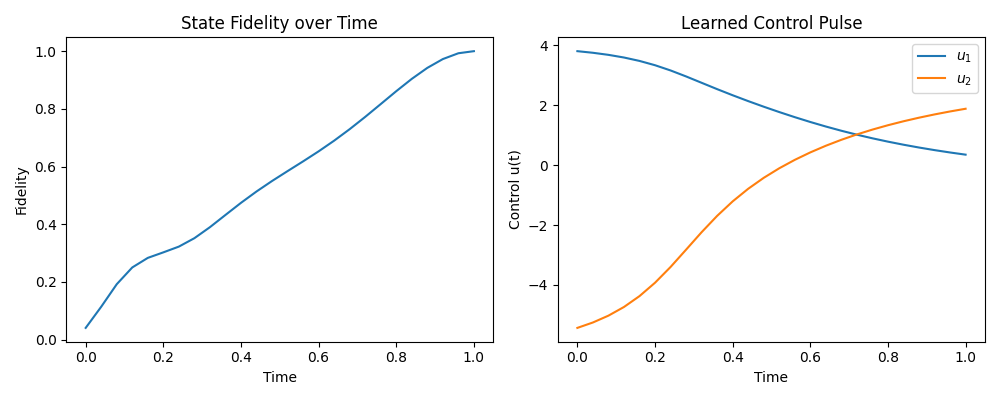}
    \caption{Fidelity evolution over time and the control amplitudes $u_1$ and $u_2$.}
    \label{fig:steerable_control}
\end{figure}

Given a source state and a target state, our goal is to construct a parameterized quantum circuit that acts through controlled Hamiltonian dynamics to steer the source towards the target.
In that, we follow these steps:
\begin{enumerate}
    \item
    For $n$ equals 2, 3, or 4 qubits, {we consider the \emph{Heisenberg model} as the drift Hamiltonian and the three \emph{Pauli operators} as controls. Namely, we set} 
    \begin{equation}
    \begin{cases}
    H_0  &= \sum_{k=1}^{n-1} X_{k} X_{k+1} + Y_k Y_{k+1} + Z_k Z_{k+1}   \quad \text{ and } \\ 
    H_k &= 
    \begin{cases}
    X_k &\text{ if $k \equiv 1 \mod 3$,}
    \\ 
    Y_k &\text{ if $k \equiv 2 \mod 3$,}
    \\ 
    Z_k &\text{ if $k \equiv 0 \mod 3$,}
    \end{cases} \quad \text{ for $1\leq k \leq n$.}
    \end{cases}
    \label{eq:control_hamiltonians}
    \end{equation}
    
    \item 
    Let $|p_0\rangle$ and $|p_1 \rangle$ be the color encodings of two selected regions of the canvas.
    Details on the color encoding are provided at the end of the section.
    We consider the time optimal control problem defined by equations \eqref{eq:opt_control} and \eqref{eq:QNODE}.

    Due to the choice of Hamiltonians in~\eqref{eq:control_hamiltonians},
    the parametrized quantum circuit $U(1)$ defined in \eqref{eq:splitting circuit} can be written as a sequence of rotation gates $R_X, R_Y, R_Z$, which are provided in standard quantum utilities such as \texttt{qiskit}.
    Namely, we have
    \begin{equation}\label{eq:splitting circuit_rotation_gate}
    \begin{aligned}
    U(1) \approx  &\prod_{k=1}^N \left ( \exp(-i H_0 \frac{\Delta t}{2}) R_{X_1}\left(- u_1(k\Delta t) \frac{\Delta t}{4}\right) R_{Y_2}\left (- u_2(k\Delta t) \frac{\Delta t}{4}\right ) \ldots
    \right.
    \\
    &\quad \quad
    \left . R_{Y_2}\left (- u_2(k\Delta t) \frac{\Delta t}{4}\right )
    R_{X_1}\left(- u_1(k\Delta t) \frac{\Delta t}{4}\right) \exp(-i H_0 \frac{\Delta t}{2})
    \right )
    \end{aligned}
    \end{equation}
    where $N$ is a user-defined smoothness parameter, and $\Delta t = 1/N$.
    \begin{figure}[H]
    \centering
    \includegraphics[width=\textwidth]{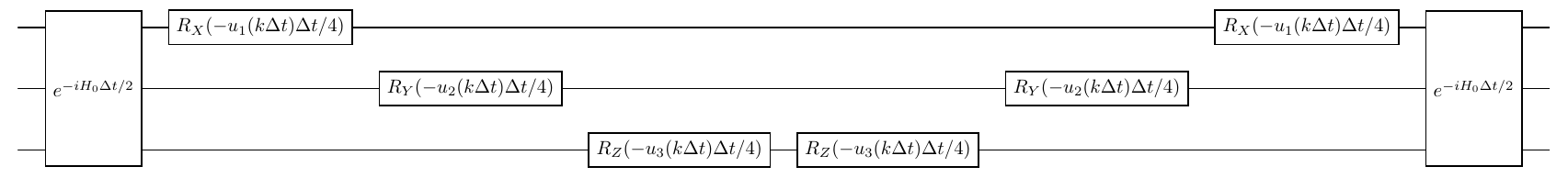}
    \caption{The $k$-th part of the splitting circuit \eqref{eq:splitting circuit_rotation_gate} for 3 qubits}
    \label{fig:splitting_circuit_tikz}
    \end{figure}

    \item We train the neural network $\boldsymbol{u} (t)$ by minimizing \eqref{eq:opt_control}. 
    Forward evaluation is done using \eqref{eq:splitting circuit_rotation_gate} in PennyLane, and we backpropagate gradients through the neural network using JAX.
\end{enumerate}
Once training is complete, we evaluate $|\rho(t)\rangle$ in the same way as in step 2 for the user-defined time $t>0$, and then paste the resulting state into the target region. 
A simulated trajectory that illustrates the steering of one state to another in the 2-qubit state space is shown in Figures \ref{fig:steerable_NN} and \ref{fig:steerable_control}.

Finally, we elaborate on how we obtain source and target states from selected regions.
Encoding every pixel into a qubit can become computationally expensive for high-resolution images. 
Inspired by the Collage brush~\cite{ferreira2025quantumbrush}, we reduce the color information using Singular Value Decomposition (SVD).
Specifically, let $C$ be the $m\times d$ matrix encoding (\(d=3\) for RGB and \(d=4\) for RGBA) of \(m\) pixels. 
Its SVD is given by
\[
C = U S V,
\]
where \(U\) is semi-unitary, \(S \) is a $d\times d$ diagonal with strictly positive entries, and \(V\) is unitary.
We note that, while Collage uses RGB encoding, we use RGBA, where the extra "A" is a parameter for transparency (often denoted alpha in computer vision).

In our implementation, we provide the following three options:
\begin{itemize}
    \item \(2\) qubits: We use the logarithm of the \(4\) values on the diagonal of \(S\).
    We denote it as \(\hat{S} = \log (S) = \{\log(s_i)|s_i\in \texttt{diag}(S)\}\);
    \item \(3\) qubits: We consider \(4\) values of \(\hat{S}\) and \(4\) additional values from \(V \hat{S}\);
    \item \(4\) qubits: use the \(16\) coefficients from four vectors of \(\hat{S}\), \(V \hat{S}\), \(V^2\hat{S}\) and \(V^3 \hat{S}\).
\end{itemize}

For every number of qubits $n$, we construct the encoding as 
\begin{equation*}
\mathrm{state}  =
\begin{pmatrix}
\mathrm{Id}_d & O & O & O
\\
O & V & O & O
\\ 
O & O & \ddots & O
\\ 
O &  \cdots & \cdots & V^{2^{n-2}-1}
\end{pmatrix}
\begin{pmatrix}
\hat{S} \\ 
\hat{S} \\ 
\vdots
\\
\hat{S}
\end{pmatrix}
\text{ where }
\hat{S} = \log(S).
\end{equation*}

In the evaluation stage, the output state has the same length as the input (either the source or a newly defined paste region).
We recover these states as coefficients in the matrices \(S\) and \(V\), taking care to account for the normalization factors.
Note that the reconstructed pixels obtained by multiplying the new matrices \(U\), \(S\), and \(V\) may contain negative values or exceed \(255\). 
We simply clip the results to the valid range \([0,255]\).

\subsection{Outcome}\label{sec: outcome steerable}
\begin{wrapfigure}{R}{0.3\textwidth}
\centering
\includegraphics[width=0.3\textwidth]{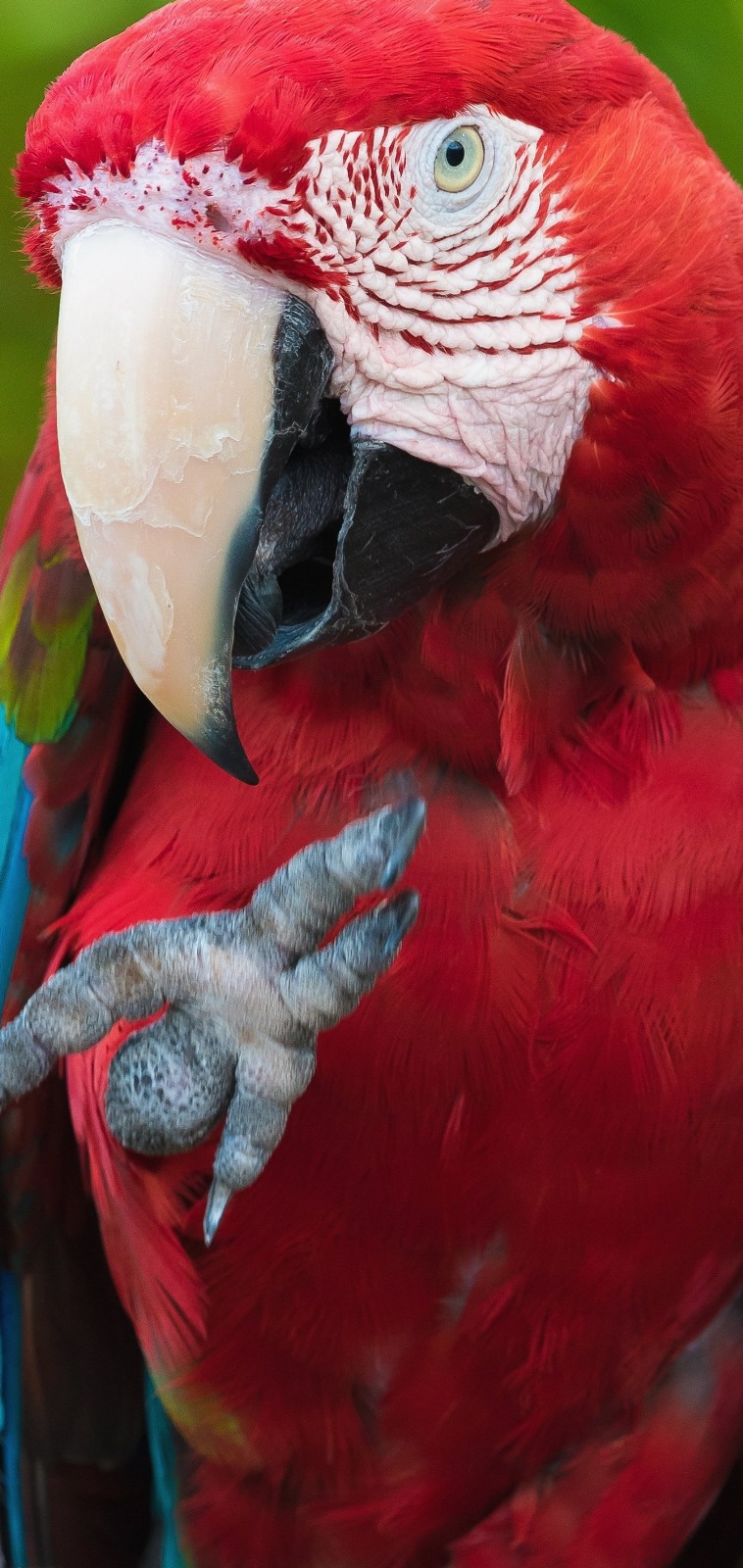}
\caption{\label{fig: parrot}Target for the application of the Steerable brush for generating Figure~\ref{fig: renoir}.}
\end{wrapfigure}

Figure~\ref{fig: renoir} shows the result of applying the Steerable effect with target Figure~\ref{fig: parrot}, taking Renoir's \emph{Bal du moulin de la Galette} \cite{renoir_bal_du_moulin_1876}.
There, we show equally spaced values of $t$ ranging from 0 to 1. 
As $t$ increases, the initially smooth blend of colors, characteristic of Renoir's work, converges toward the vivid hues of the parrot. 
Interestingly, however, when $t$ approaches 1, the colors seem to decay towards a gray spectrum.
This pattern is frequent in several of our extrapolation experiments, although not in all.
We are still unsure of the causes of this gray decay, but we believe that this might be related to the choice of Hamiltonians in equation \eqref{eq:control_hamiltonians} or to clipping the state measurements to the range \([0,255]\).
Note that the convergence to a different color palette at $t=1$ is not unexpected: perfect convergence to the target would only happen in the perfect scenario in which the final fidelity in equation~\eqref{eq:opt_control} is maximized. 
Unfortunately, for practical training, this will hardly be the case, meaning that the expected convergence to the target colors at the $t=1$ limit will likely only be approximated. 
Adding more controllability or experimenting with different approximation algorithms and color encoding might bring the state when $t=1$ closer to the target.

\begin{figure}[!t]
    \centering
    % \foreach \file in {
    % fig/andy/Monroe_0_to_1_c2_t0p00.png,
    % fig/andy/Monroe_0_to_1_c2_t0p14.png,
    % fig/andy/Monroe_0_to_1_c2_t0p29.png,
    % fig/andy/Monroe_0_to_1_c2_t0p43.png,
    % fig/andy/Monroe_0_to_1_c2_t0p57.png,
    % fig/andy/Monroe_0_to_1_c2_t0p71.png,
    % fig/andy/Monroe_0_to_1_c2_t0p86.png,
    % fig/andy/Monroe_0_to_1_c2_t1p00.png,
    % fig/andy/Monroe_1_to_0_c2_t0p00.png,
    % fig/andy/Monroe_1_to_0_c2_t0p14.png,
    % fig/andy/Monroe_1_to_0_c2_t0p29.png,
    % fig/andy/Monroe_1_to_0_c2_t0p43.png,
    % fig/andy/Monroe_1_to_0_c2_t0p57.png,
    % fig/andy/Monroe_1_to_0_c2_t0p71.png,
    % fig/andy/Monroe_1_to_0_c2_t0p86.png,
    % fig/andy/Monroe_1_to_0_c2_t1p00.png
    % }{%
    %     \includegraphics[width=0.22\linewidth]{\file}
    % }
    \includegraphics[width=0.22\linewidth]{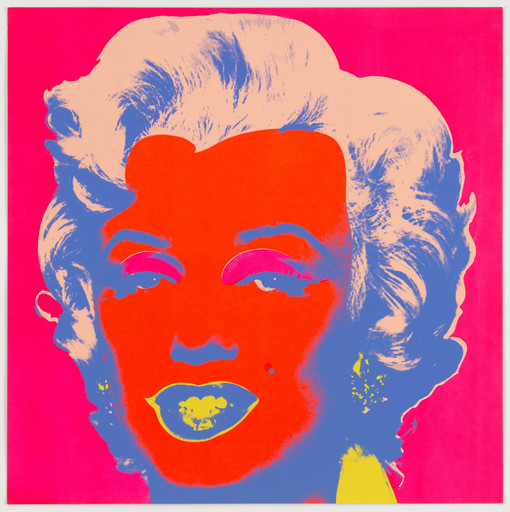}
    \includegraphics[width=0.22\linewidth]{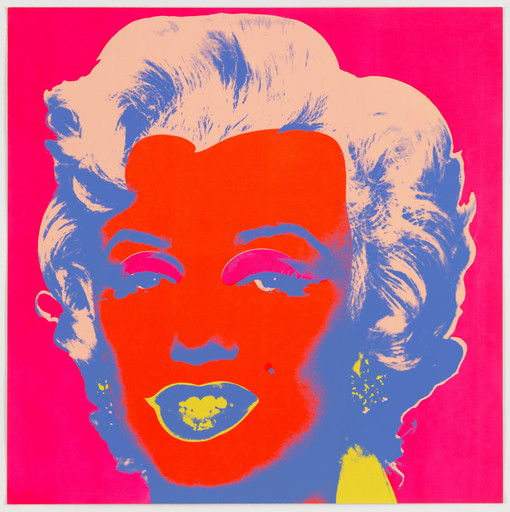}
    \includegraphics[width=0.22\linewidth]{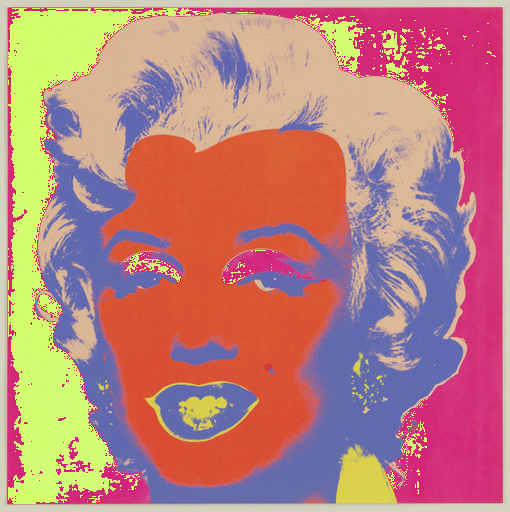}
    \includegraphics[width=0.22\linewidth]{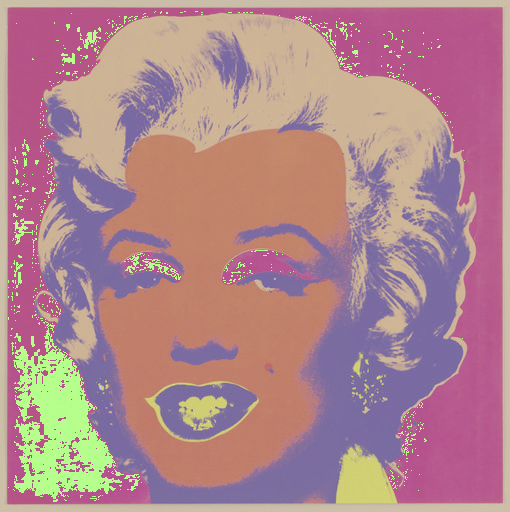}
    \includegraphics[width=0.22\linewidth]{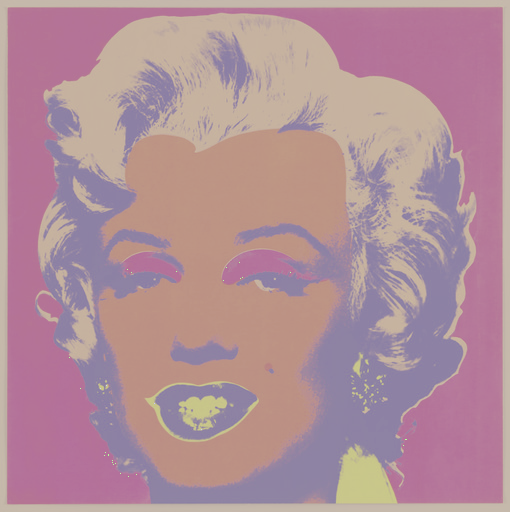}
    \includegraphics[width=0.22\linewidth]{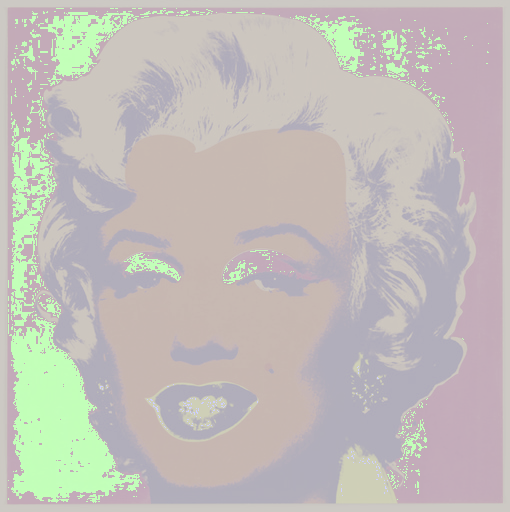}
    \includegraphics[width=0.22\linewidth]{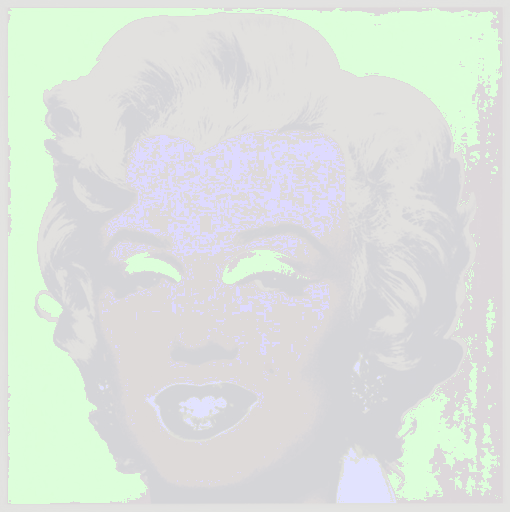}
    \includegraphics[width=0.22\linewidth]{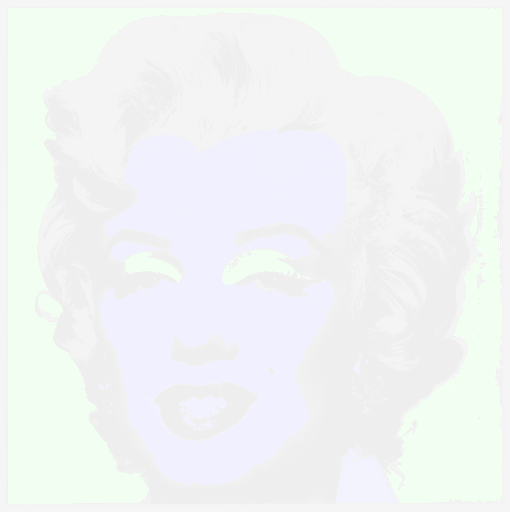}
    \includegraphics[width=0.22\linewidth]{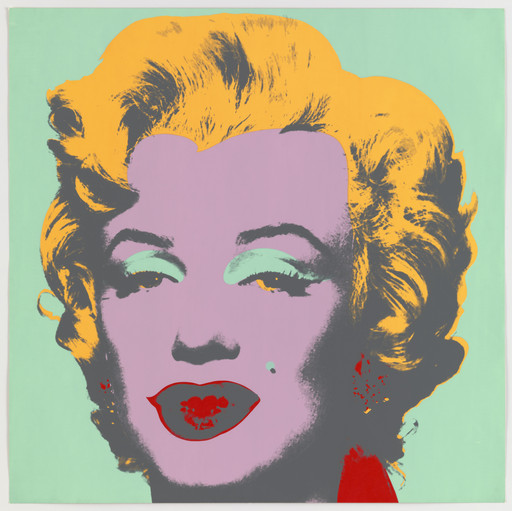}
    \includegraphics[width=0.22\linewidth]{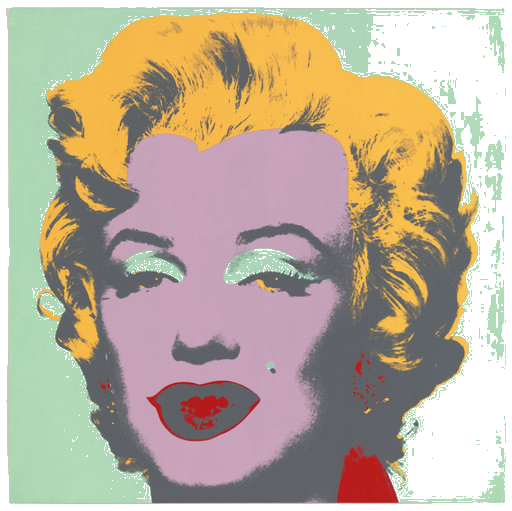}
    \includegraphics[width=0.22\linewidth]{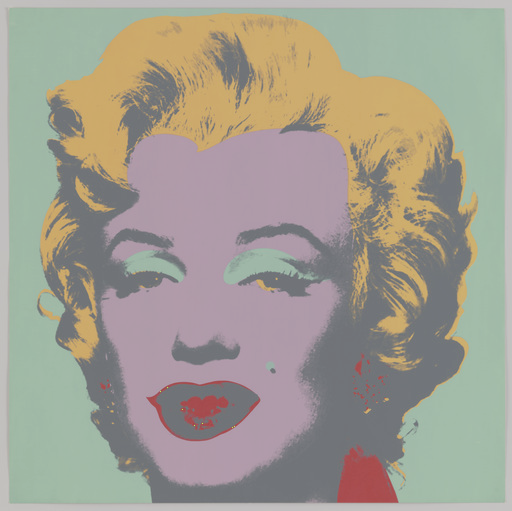}
    \includegraphics[width=0.22\linewidth]{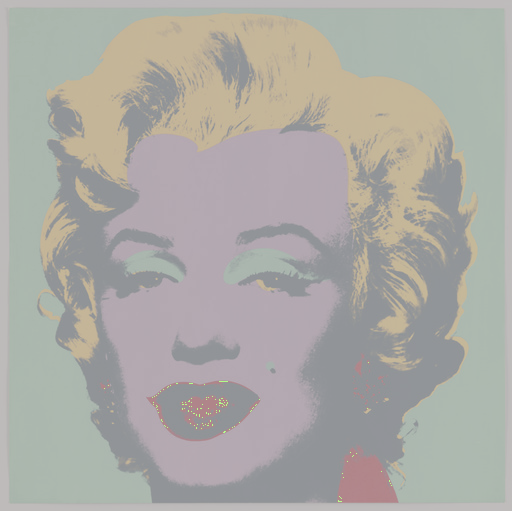}
    \includegraphics[width=0.22\linewidth]{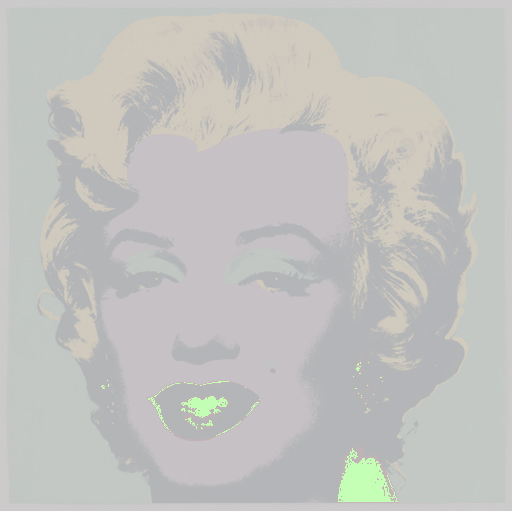}
    \includegraphics[width=0.22\linewidth]{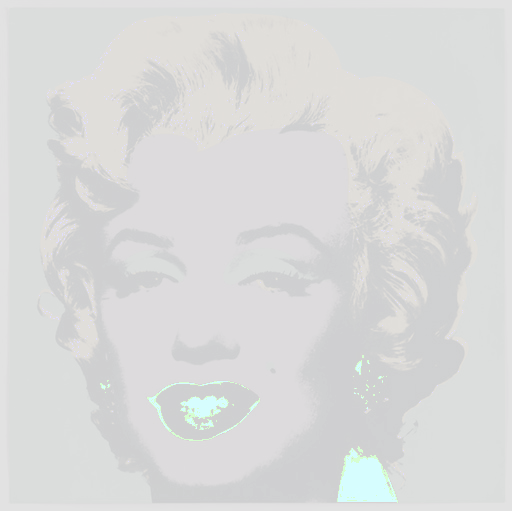}
    \includegraphics[width=0.22\linewidth]{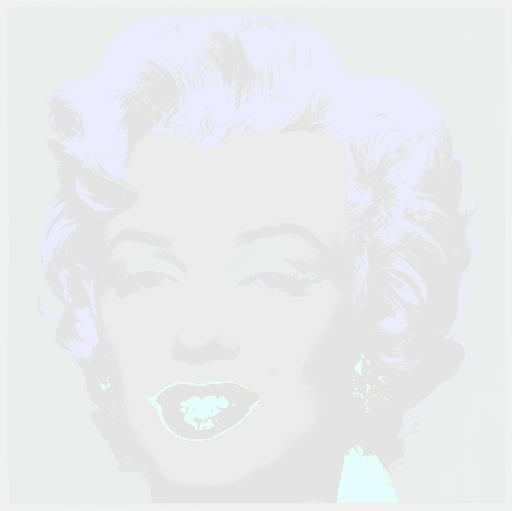}
    \includegraphics[width=0.22\linewidth]{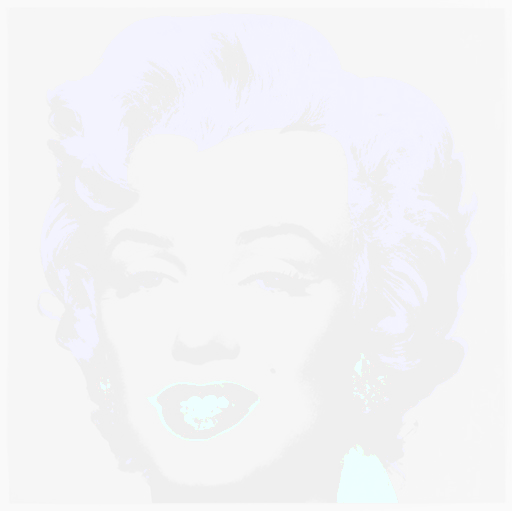}
    \caption{Steerable effect applied to Warhol's \emph{Marilyn Diptych}. The leftmost images of the first and third rows were used as the source and target.}\label{fig: andy}
\end{figure}

Figure~\ref{fig: andy} shows the application of Steerable with \texttt{controls} equal to 2 on two images of Andy Warhol's \emph{Marilyn Diptych} silk screen series \cite{warhol1962marilyn} (recall that \texttt{controls} also corresponds to the number of qubits in the quantum circuit).
The first two rows indicate applications of Steerable for $t$ that vary uniformly from 0 to 1, where the target is the leftmost image of the third row.
Similarly, the last two rows indicate applications of steerable targeting the leftmost image of the first row.
Note that while the use of Steerable depicted on the first two rows drifts the very reddish colors of the source to the correct green and blue directions, it fails to converge the colors of the second source towards red in the last couple of rows.
Moreover, in both cases, the aforementioned convergence towards gray can also be observed.
In our experiments with Warhol's work, we have noticed that this tendency towards gray is significantly mitigated by introducing more control; however, the colors for mid-range values of $t$ start blending rapidly, which directly jeopardizes visualization.

\begin{figure}[t!]
    \centering
    \begin{subfigure}{0.48\textwidth}
        \centering
        \includegraphics[width=\linewidth]{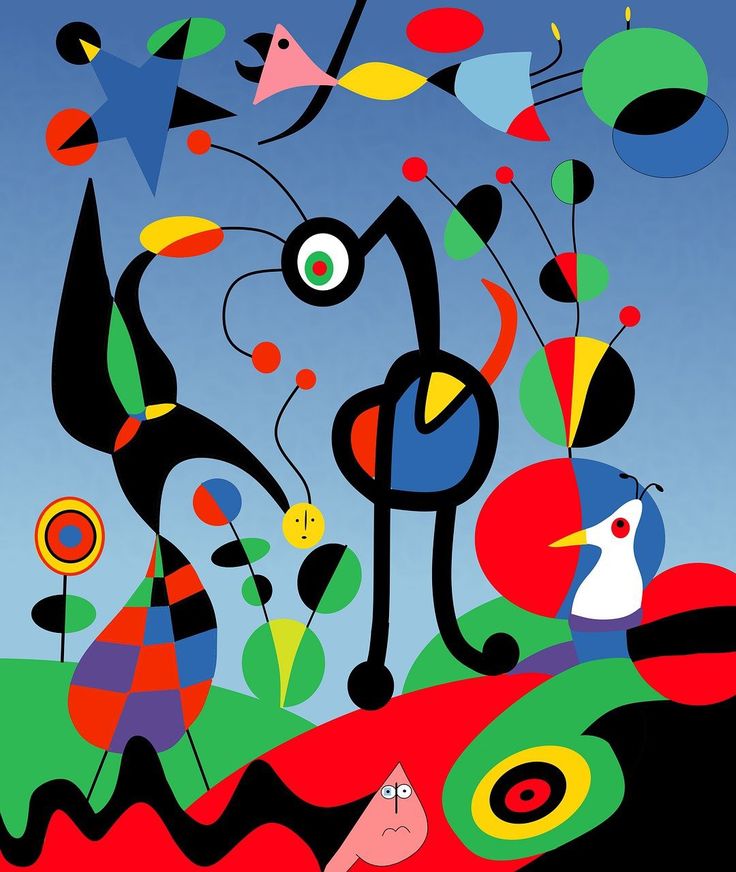}
        \caption{Original \emph{El Jardín} by Joan Miró.}
    \end{subfigure}\hfill
    \begin{subfigure}{0.48\textwidth}
        \centering
        \includegraphics[width=\linewidth]{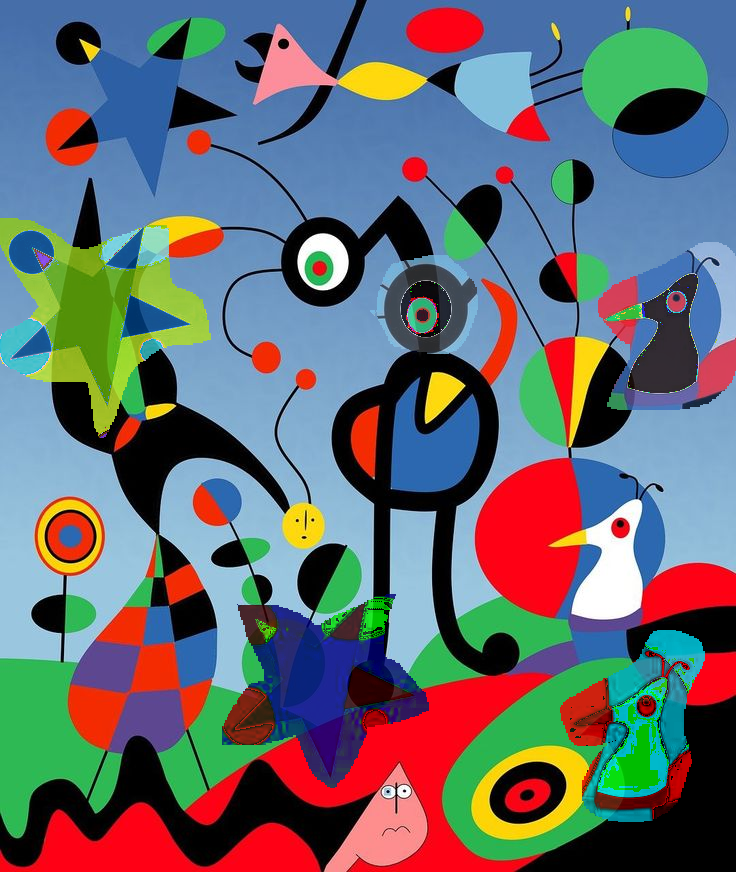}
        \caption{Application of Steerable to Miró's piece.}
    \end{subfigure}
    \caption{}  \label{fig: miro}
\end{figure}

Finally, Figure~\ref{fig: miro} shows an application of the steerable brush in Joan Miró’s \emph{El Jardín} \cite{Miro1925_ElJardin}. 
In particular, we select the white bird as the source and steer it into two additional bird shapes using 2 (top) and 3 (bottom) controls. The star in the top-left corner is likewise steered into two copies using 3 controls (middle) and 4 controls (bottom left). 
The parameter $t$ was set to values in between 0.2 and 0.4.
We also paste several other small elements, which we leave to the reader as a playful exercise of "Where’s Charlie?".

\section{Chemical}\label{sec: chemical}
%\subsection{Motivation}
In its century of existence, quantum mechanics has been responsible for several technological and scientific revolutions. 
Still, few were as groundbreaking as the description of the periodic table through Pauli's exclusion principle.
The scientific understanding that atomic particles behave differently from the old Newtonian rules of physics forever entangled the fields of quantum mechanics and chemistry, making the study of the former a foremost ingredient in the latter.
For instance, quantum mechanical considerations imply that molecules -- amalgamations of atoms sharing electrons -- are only allowed to be found within some discretely separated energetic states.
Being a little more explicit here, a hydrogen molecule, $H_2$, for example, is associated with a whole tower of energy levels $E_0, E_1, E_2,\dots$, but it can never be found with energy equal to a value between two of these numbers.
In other words, molecular energy is \emph{quantized}.
For the most part, molecules are found in the minimum energy state possible, $E_0$, their \emph{ground state}, although sometimes they will be seen in higher states called \emph{excitations}.
As it turns out, finding the ground states of molecules is an extremely complicated computational problem, and techniques based on quantum computers have become competitive candidates to deal with it.

VQE is one of these techniques.
It expresses the energy of a molecule (or, to be precise, its expectation value) as an objective function in an optimization program, and, at each step, it searches for circuits that minimize this quantity, which is equivalent to saying that it searches for the molecule's ground state energy.
When it works well -- which unfortunately is not the case for every single molecule -- VQE finds not only $E_0$, but also a last outputted circuit that transforms a reference state into the molecule's ground state.

\emph{Chemical} serves as a visualization of VQE. 
It uses the algorithm's parametric circuits to drive changes at each pixel crossed by the stroke.
Once more, when the algorithm correctly converges, its last transformation corresponds to the evolution of a molecule towards the ground state.
In this sense, the brush encodes not only the well-known VQE algorithm, highlighting the most modern advances of quantum research, but also incorporates Nature itself into the process.

\subsection{Theoretical framework}\label{sec: theory chemical}
In the VQE formulation of variational algorithms, we suppose that $H$ is a Hermitian operator corresponding to the solution of Schrödinger's equations for some molecules\footnote{Hands-on introductions and review of VQE can be found in~\cite{fedorov2022vqe, IBMQuantum_VQE_2025}. For a more in-depth theoretical review, refer to~\cite{tilly2022variational}.}.
The exact value of $H$ depends not only on the type of molecule, but also on some physical constants such as Planck's and the average distance between the atoms in the molecule -- also known as the molecule's \emph{bond distance}.
By the Spectral Theorem~\cite{hall2013quantum}, $H$ can be written as 
\[
H = \sum_{k=0}^{d-1} E_k \, \ket{\phi_k}\!\bra{\phi_k}.
\]
The eigenstates $\{\ket{\phi_k}\}_{k=0}^{d-1}$ represent the allowed configurations of the electrons in the molecule, while the quantities $E_k$ are the associated energies.
In particular, we may always assume $E_0\leq E_1\leq \dots\leq E_{d-1}$, in which case the minimum energy allowed, $E_0$, is the ground state of the molecule.

The variational principle implies that, given any other state $\ket{\psi}$ (not necessarily equal to one of the eigenstates $\ket{\phi_k}$, but potentially a linear combination of those), the quantity $\bra{\psi}H\ket{\psi}$ is at least as big as $E_0$, with equality if and only if $\ket{\psi}$ is proportional to $\ket{\phi_0}$.
Therefore, the problem of minimizing the quantity $\bra{\psi}H\ket{\psi}$ is fully equivalent to finding the ground state of $H$, up to a proportionality factor.
In practice, we look for a state $\ket{\psi}\neq 0$ such that
\begin{equation}\label{eq: program}
E^\star=\min R(\psi)\ge E_0,
\end{equation}
where
\[
R(\psi)=\frac{\langle \psi|H|\psi\rangle}{\langle \psi|\psi\rangle}
\]
is known as the \emph{Rayleigh-Ritz} quotient.

The optimization program for equation \eqref{eq: program} is the part usually tackled by VQE.
This is possible by mapping the states of the molecule's electrons into the qubits of a quantum circuit and the Hamiltonian into a linear combination of Pauli gates\footnote{We here assume the Jordan–Wigner transform \cite{jordan1928pauli} as such a map. A slightly more popular alternative is the Bravyi-Kitaev transform~\cite{bravyi2002fermionic}, whose pros and cons are reviewed in~\cite{tranter2015b}. We point out that these maps depend on the choice of a basis as a hyperparameter; for all experiments, we choose the basis to be $\text{STO-3G}$.}.
Then, a parametrized family of circuits $U(\boldsymbol{\eta})$ is applied to a reference state $\ket{\Phi_{\text{HF}}}$, called the \emph{Hartree-Fock reference state}, and the values of $\boldsymbol{\eta}$ are adjusted according to some classical continuous optimization routine -- such as line-search or fixed-step gradient descent -- to minimize
\begin{equation*}
E(\boldsymbol{\eta})=\bra{\psi(\boldsymbol{\eta})}H\ket{\psi(\boldsymbol{\eta})},
\end{equation*}
where 
\[
\ket{\psi(\boldsymbol{\eta})}=U(\boldsymbol{\eta})\ket{\Phi_{\textbf{HF}}}.
\]

The exact form of the parametric circuits $U(\boldsymbol{\eta})$ corresponds to the ansatze of the solution and can be taken to better correspond to the chemical system or to guarantee efficiency of the hardware.
Here, we opt for the more chemically motivated disentangled unitary coupled cluster ansatz (DUCC)\footnote{However, the choice of DUCC as an ansatz introduces a second hyperparameter in the form of the ordering of the unitary operators, which, in general, do not commute. 
Different orderings may lead to different convergence trajectories to the ground state and, consequently, to different parameters for the circuits.
Here, we take the so-called standard ordering of operators, although we stress that different choices, including usually faster converging ones, are possible.}.

In each optimization step $i$ in which $\boldsymbol{\eta}_i$ is updated, there will be associated $f_i=f(\ket{\psi(\boldsymbol{\eta}_i)})$ and $U_i=U(\boldsymbol{\eta}_i)$.
In particular, provided that there is good convergence to $E^\star$, we might assume that $\{f_i\}_{i=1}^N$ is an overall decreasing sequence and that $\{U_i\}_{i=1}^N$ gets closer and closer to the process that maps the reference state to the minimizer in equation \eqref{eq: program} -- i.e., $\ket{\psi(\boldsymbol{\eta}^\star)}$ becomes a good approximation of the ground state. 
The parametric list of circuits $U(\boldsymbol{\eta}_i)$ is the main quantum ingredient of the brush's implementation.

\subsection{Implementation}\label{sec: chemical implementation}
\begin{figure}[t!]
   \centering
    \includegraphics[width=0.6\textwidth]{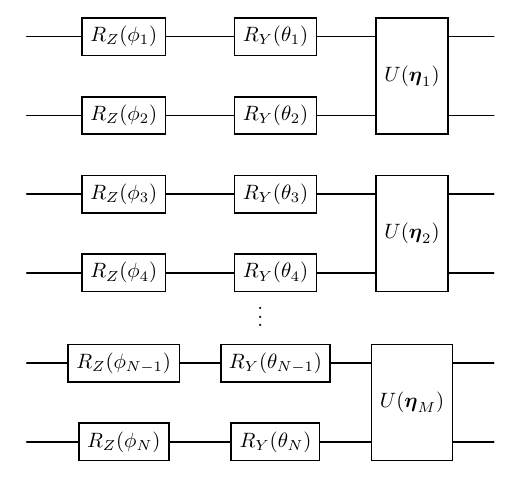}
    \caption{Basic circuit of Chemical. Measurements are not shown. Note that here, $n_{qub}$ equals 2.}
    \label{fig: circuit}
\end{figure}

The user begins by drawing one or more strokes of some fixed radius on the canvas.
The strokes define arrays of pixels, whose hue and luminosity values (HL) are aggregated by averaging the angular HL of neighboring pixels\footnote{Refer to the appendix of \cite{ferreira2025quantumbrush} for in-depth explanations of these operations.}.
Each of these aggregated HL values $\{(\phi_j,\theta_j)\}_{j=1}^N$ is encoded as qubits through applications of $R_Z(\phi_j)$ and $R_Y(\theta_i)$ gates (see Figure~\ref{fig: circuit}).
The number of qubits $N$ is chosen to match the total number of pixels in the whole estimated parametric family \{$U(\boldsymbol{\eta}_i)\}_{i=1}^M$; recall that each circuit of the family admits the same number of qubits $n$, meaning $N=M\times n$.
For example, in our choices of hyperparameters, $n$ for the $H_2$ molecule equals $4$.
We follow \cite{ferreira2025quantumbrush} and denote the thus encoded states by $\ket{(\theta_j,\phi_j)}$.

Next, $U(\boldsymbol{\eta}_i)$ is applied to each $\ket{(\theta_j,\phi_j)}$, ultimately changing the encoded states according to the VQE algorithm.
The new HL values are reconstructed from the trigonometric transformations of the expectations of the Pauli operators, as described in the appendix of \cite{ferreira2025quantumbrush}.
Once again, assuming convergence of the VQE, the qubits at the end of the stroke will be mapped to the state $U(\boldsymbol{\eta}^\star)\cdot\bigotimes_{j=N-n_{qub}}^N\ket{(\phi_j,\theta_j)}$, where $U(\boldsymbol{\eta}^\star)$ is a good approximation of the operator that maps $\ket{\Phi_{\text{HF}}}$ to the ground states.
Of course, $\ket{\Phi_{\text{HF}}}$ is likely very different from $ \bigotimes_{j=N-n_{qub}}^N\ket{(\phi_j,\theta_j)}$ and, consequently, $U(\boldsymbol{\eta}^\star)\cdot \bigotimes_{j=N-n_{qub}}^N\ket{(\phi_j,\theta_j)}$ will not represent the molecule's minimum energy state.
This is where the beauty of Schrördinger's equation lies: $U(\boldsymbol{\eta}^\star)$ is an approximation of the \emph{physical process} that takes the Hartree-Fock state to the ground, but this same process can also be applied to any initial quantum state, including an encoding of HL values.
In a sense, we are not interested here in the output of the VQE algorithm, but rather in the output process of the algorithm.

The user of Chemical can choose as a hyperparameter \texttt{bond distance} used by the VQE.
Large bound distances are expected to lead towards faster convergence as the molecular energy becomes well-approximated by two distant hydrogen atoms.
Another parameter set by the user is the \texttt{number of repetitions}, which varies from 0 to 100 and controls the number of times that the circuits in the parametrized family are repeatedly applied to neighboring qubits.
Its object is to make the transition between different circuit applications smoother, which is particularly significant for monochromatic pieces. 
Finally, the stroke's \texttt{radius} is the last parameter that is adjusted by the user; it defines the thickness of the stroke.

Two subtleties about Chemical's implementation should be highlighted here. 
First, as of now, $H_2$ is the only molecule available. 
This is due to the simplicity of the VQE algorithm when applied to it.
For the hyperparameters we chose, VQE converges pretty fast to the ground state of $H_2$, which can be understood as a manageable number of circuits $M$.
For other molecules, these hyperparameters lead to very slow convergences, which implies values of $M$ some orders of magnitude above the number of qubits created by a usual stroke.
Therefore, implementing Chemical for more molecules would require fine-tuning of the hyperparameters, which is an area of research of its own.

\begin{figure}[t!]
    \centering
    \begin{subfigure}{0.48\textwidth}
        \centering
        \includegraphics[width=\linewidth]{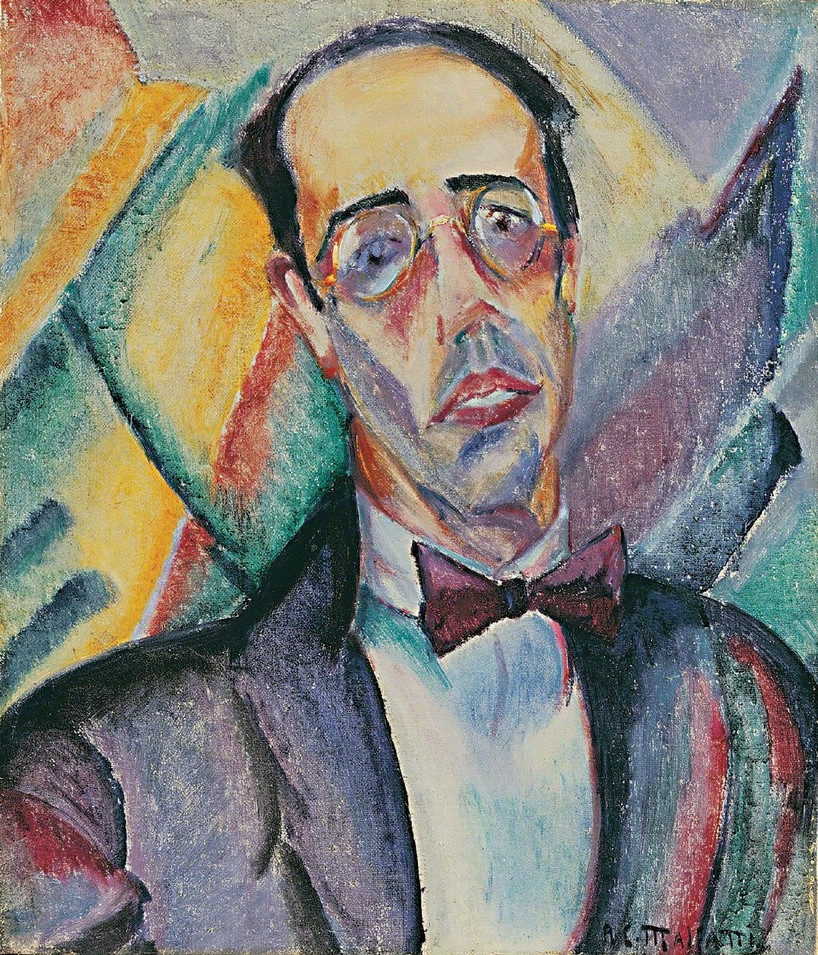}
        \caption{Original \emph{Retrato de Mário de Andrade} by Anita Malfatti.}
    \end{subfigure}\hfill
    \begin{subfigure}{0.48\textwidth}
        \centering
        \includegraphics[width=\linewidth]{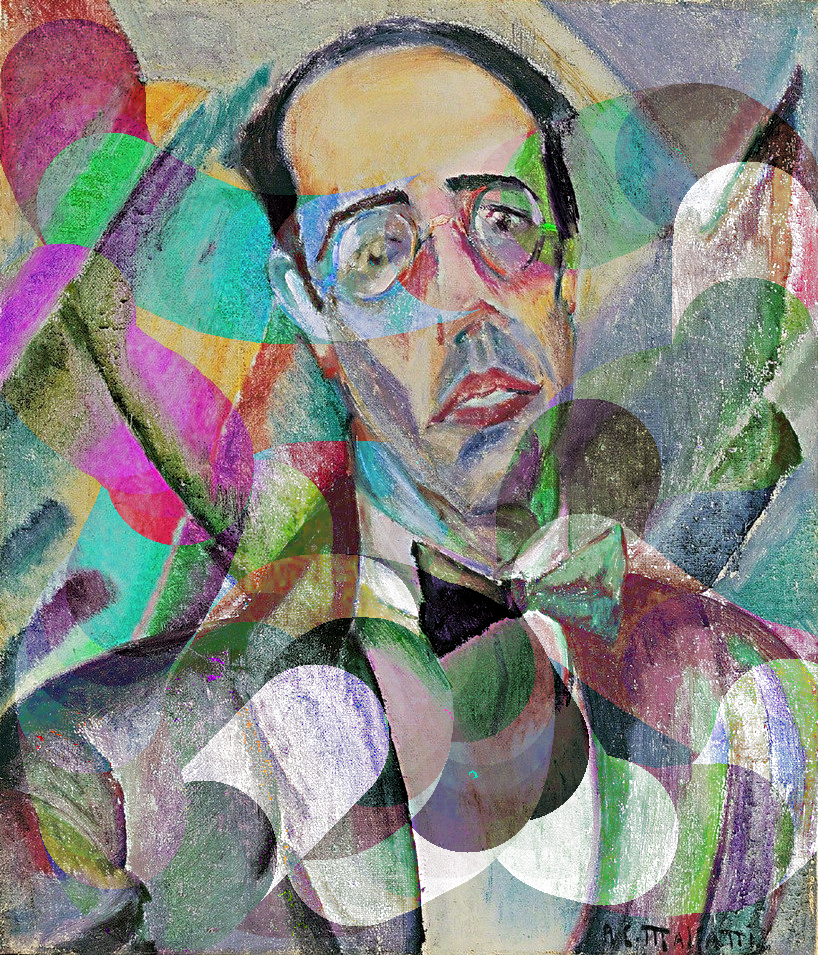}
        \caption{Application of Chemical to Malfatti's original piece.}
    \end{subfigure}
    \caption{}\label{fig: anitta}
\end{figure}

Second, instead of running VQE on the user's machine at each application of the brush, we preprocessed the parametrized families $\{U(\boldsymbol{\eta}_i)\}_{i=1}^M$ at different distance values, which we save as JSON files in a data folder. 
Although VQE tends to be fast for $H_2$, we believe that taking the bulk of the computation off the user's end improves the experience.
Nevertheless, this restricts the applications of Chemical by making it so that only a finite, despite large, set of bond distances can be used in practice.
We precomputed VQE families for 1000 bond distances, varying uniformly in the interval from 0.725 to 2.5 \r{A}. 
The user can choose any floating-point number within this range, but when applying the effect, their choice is projected onto the closest distance for which circuits were precomputed.

\subsection{Outcome}\label{sec: outcome chemical}

\begin{figure}[t!]
    \centering

    % Top picture
    \begin{subfigure}{0.7\textwidth}
        \centering
        \includegraphics[width=\textwidth]{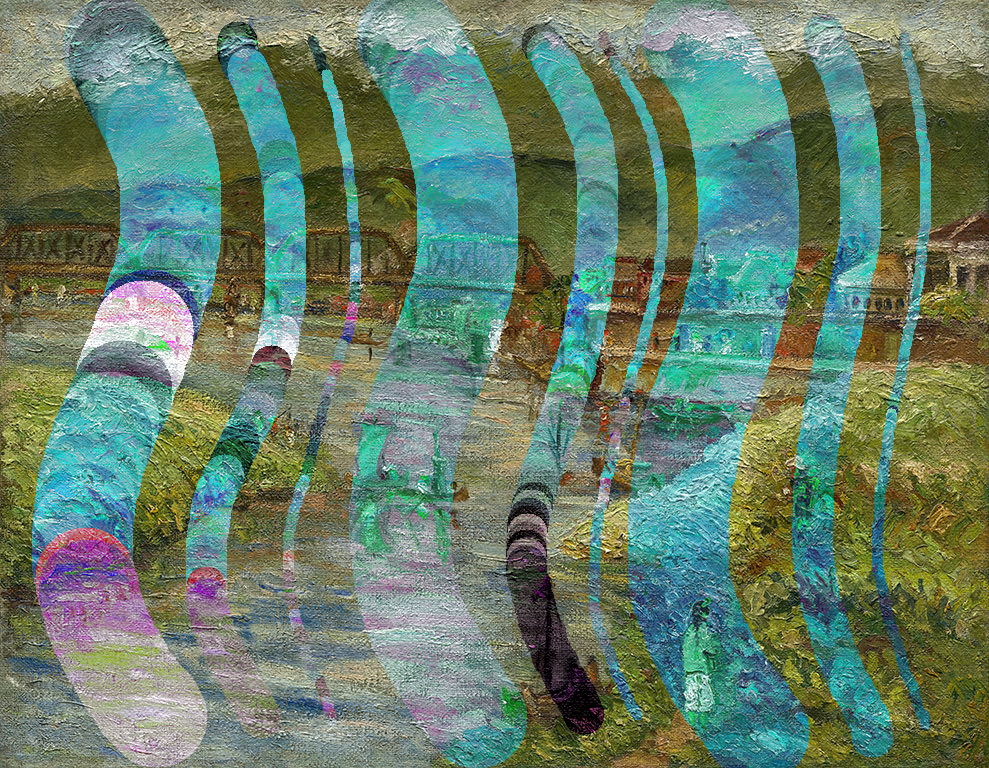} %Tân Têng-pho : Taipei Bridge (1933)
        \caption{Application of Chemical with different stroke thicknesses and bond distances to \emph{Taipei Bridge, Tamsui River} by Tân Têng-pho.}
        \label{fig:taipei_bridge}
        %\label{fig:fried_rice}
    \end{subfigure}

    \vspace{1em} % optional spacing

    % Bottom picture
    \begin{subfigure}{0.7\textwidth}
        \centering
    \includegraphics[width=\textwidth]{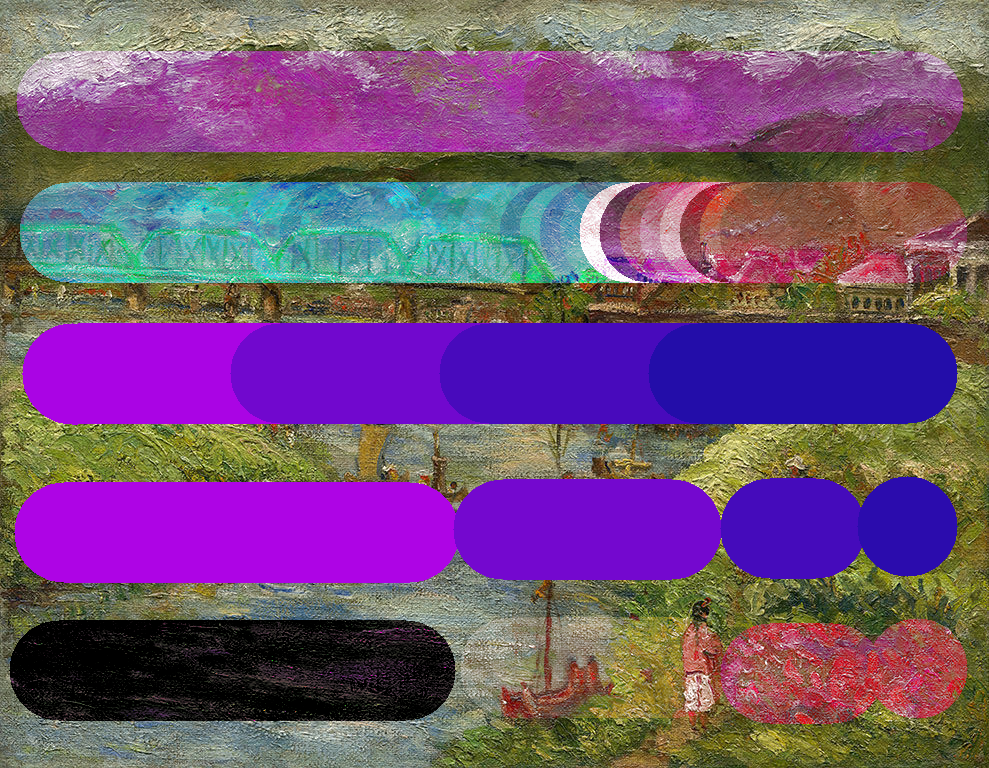} %Tân Têng-pho : Taipei Bridge (1933)
        \caption{Brush effect comparison (from top to bottom):
    Aquarela,
    Chemical,
    Heisenbrush (continuous),
    Heisenbrush (discrete),
    and Smudge.}
    % \label{fig:brush_compare}
\label{fig:taipei_bridge_compare}
    \end{subfigure}
    \caption{}
\end{figure}

We apply Chemical to Anita Malfatti's portrait of the Brazilian author Mário de Andrade \cite{Malfatti1921_PortraitMarioDeAndrade} in Figure~\ref{fig: anitta}. 
The effect here is to highlight Anita's strong blend of colors, smoothening the already remarkable mixtures of her work.
A large radius was chosen, and different values for the bond distance were experimented. 
In the end, we opted for 0.735 \r{A}.
We note that the disks generated by the averaging of HL values are clearly visible in the canvas.
As mentioned in Section \ref{sec: chemical implementation}, although changing the number of repetitions usually mitigates this effect for monochromatic works, for a strong mixture of shades as Malfatti's portrait, this is not usually the case.

In Figure~\ref{fig:taipei_bridge}, it is possible to see the application of Chemical with different stroke thicknesses and bond distances to Tân Têng-pho's \emph{Taipei Bridge, Tamsui River} \cite{ChenChengPo_1933_TaipeiBridge}.
The distance parameter was set to 0.735 \r{A} for the three leftmost strokes, 1.6 \r{A} for the middle three strokes, and 2.5 \r{A} for the three right strokes.
We see that, at least for $H_2$, the smaller the bond distance, the more variety in the stroke colors, although it is hard to make this a general rule-of-thumb, as it might be a feature of the particular convergence rate of VQE only.
Moreover, it is interesting to see that although the strokes start at the same state (top of the figure), modifying the distance seems to imply changing the estimated ground state, which is naturally expected given the Hamiltonian dependence on the average bond distances.

Figure~\ref{fig:taipei_bridge_compare} compares the effect of Chemical with other quantum brushes from~\cite{ferreira2025quantumbrush}, namely Aquarela, Heisenbrush (continuous and discrete), and Smudge.
Heisenbrush does not take into account the underlying color of the canvas. The other two brushes (Aquarela and Smudge) rely on more similar mechanisms.
All these methods transform local color information, but they do so using different quantum circuits. 
Aquarela depends on a user-defined color (purple in Figure~\ref{fig:taipei_bridge_compare}), whereas Smudge uses a predefined quantum circuit that mimics a damping effect. 
Chemical exhibits greater variability due to its evolving nature, resulting in richer and more diverse color patterns compared to the other brushes.

\bibliography{main}

\appendix

\end{document}